\documentclass[preprint,aps,showpacs]{revtex4}
\usepackage{amsmath}
\usepackage{amssymb}
\usepackage{bm}
\usepackage{graphics}
\usepackage[dvips]{graphicx,color}
\usepackage{dcolumn}% Align table columns on decimal point
\def\lesssim{\ \raise.3ex\hbox{$<$}\kern-0.8em\lower.7ex\hbox{$\sim$}\ }
\def\gesim{\ \raise.3ex\hbox{$>$}\kern-0.8em\lower.7ex\hbox{$\sim$}\ }

\newcommand{\bs}{\boldsymbol}
\newcommand{\lef}{\left}
\newcommand{\rig}{\right}
\newcommand{\up}{\uparrow}
\newcommand{\down}{\downarrow}

\begin{document}
\title{Superfluid Fermi atomic gas as a quantum simulator for the study of neutron-star equation of state}
\author{Pieter van Wyk$^1$, Hiroyuki Tajima$^2$, Daisuke Inotani$^1$, Akira Ohnishi$^3$, and Yoji Ohashi$^1$}
\affiliation{
$^1$ Department of Physics, Keio University, 3-14-1 Hiyoshi, Kohoku-ku, Yokohama 223-8522, Japan \\
$^2$ Nishina Center, RIKEN, Wako, Saitama 351-0198, Japan \\
$^3$ Yukawa Institute for Theoretical Physics, Kyoto University, Kyoto 606-8502, Japan}
\date{\today}
\begin{abstract}
We theoretically propose an idea to use an ultracold Fermi gas as a quantum simulator for the study of the neutron-star equation of state (EoS) in the low-density region. Our idea is different from the standard quantum simulator that heads for {\it perfect} replication of another system, such as a Hubbard model discussed in high-$T_{\rm c}$ cuprates. Instead, we use the {\it similarity} between two systems, and theoretically make up for the difference between them. That is, (1) we first show that the strong-coupling theory developed by Nozi\`eres-Schmitt Rink (NSR) can quantitatively explain the recent EoS experiment on a $^6$Li superfluid Fermi gas in the BCS (Bardeen-Cooper-Schrieffer)-unitary limit far below the superfluid phase transition temperature $T_{\rm c}$. This region is considered to be very similar to the low density region (crust regime) of a neutron star (where a nearly unitary $s$-wave neutron superfluid is expected). (2) We then theoretically compensate the difference that, while the effective range $r_{\rm eff}$ is negligibly small in a superfluid $^6$Li Fermi gas, it cannot be ignored ($r_{\rm eff}=2.7$ fm) in a neutron star, by extending the NSR theory to include effects of $r_{\rm eff}$. The calculated EoS when $r_{\rm eff}=2.7$ fm is shown to agree well with the previous neutron-star EoS in the low density region predicted in nuclear physics. Our idea indicates that an ultracold atomic gas may more flexibly be used as a quantum simulator for the study of other complicated quantum many-body systems, when we use, not only the experimental high tunability, but also the recent theoretical development in this field. Since it is difficult to directly observe a neutron-star interior, our idea would provide a useful approach to the exploration for this mysterious astronomical object.
\end{abstract}
%\pacs{03.75.Ss, 03.75.-b, 03.70.+k}
\maketitle
%%%%%%%%%%%%%%%%%%%%%%%%%%%%%%%%%%%%%%%%%%%%%%%%%%%%%%%%%%%%%%%%%%%%%%%%%%%%%% 
\par
\section{Introduction}
\par
In cold atom physics, the high-tunability of this system\cite{Bloch,Chin} has realized various interesting quantum phenomena. One example is the BCS (Bardeen-Cooper-Schrieffer)-BEC (Bose-Einstein condensation) crossover phenomenon in $^{40}$K\cite{Jin} and $^6$Li\cite{Zwierlein,Kinast,Bartenstein} Fermi gases, where the character of a Fermi superfluid continuously changes from the weak-coupling BCS-type to the BEC of tightly bound molecules\cite{Eagles,Leggett,NSR,Randeria,Strinati,Ohashi,Levin,Giorgini}, with increasing the strength of a pairing interaction by adjusting the threshold energy of a Feshbach resonance\cite{Chin}. Another example is a $^{87}$Rb Bose gas loaded on an optical lattice, where the superfluid-Mott insulator transition has been realized by tuning the atomic hopping parameter between lattice sites, by adjusting the height of lattice potential\cite{Greiner,Kohl,Bloch}. 
\par
The high-tunability of ultracold atomic gases has also made us expect the usage of this system as a ``quantum simulator" for the study of other complicated quantum many-body systems\cite{Bloch2}; however, this exciting attempt has not yet reached its full potential. For example, although similarity between an ultracold Fermi gas loaded on a two-dimensional optical lattice and high-$T_{\rm c}$ cuprates\cite{Dagotto} has been pointed out\cite{Demler}, the current experimental achievement is still at the $s$-wave pairing state in the case of a very shallow three-dimensional optical lattice\cite{Chin2,Miller} (which cannot be described by the Hubbard model). The recent extensive experimental efforts have enabled us to precisely measure various physical quantities in ultracold gases\cite{Jin2,Jin3,Jin4,Jin5,Navon,Sanner,Kohl2,Ku,Horikoshi,Vale}. Thus, when an ultracold atomic gas works as a quantum simulator for another system, the high-tunability, as well as these sophisticated experimental techniques, would contribute to understanding this target system. This success would also give feedback to cold atom physics, to accelerate the further development of this field.
\par
In this paper, as a promising target of a quantum simulator made of an ultracold Fermi gas, we theoretically investigate the equation of state (EoS) of a neutron star. A neutron star is much smaller than the earth (the radius $R$ of a neutron star is about 10 km.), but the mass is comparable to the sun, so that it is considered as the densest matter in our universe\cite{Shapiro}. The recent discovery of the massive neutron star PSR J1614-2230 (with a mass $M=1.97\pm0.04M_{\odot}$, where $M_{\odot}$ is the solar mass) using the so called Shapiro delay \cite{Demorest}, along with the later discovery of PSR J0348+0342 ($M=2.01\pm0.04M_{\odot}$)\cite{Antoniadis}, have spurred a heated debate about the internal structure of this mysterious star. This is because it has theoretically been predicted that hyperons should appear deep inside a neutron star where the density $n\gesim2\rho_0$ (where $\rho_0=0.16~{\rm fm}^{-3}$ is the nuclear saturation density), and that this makes it difficult to explain the existence of such a massive neutron star\cite{Demorest,Lattimer,Takatsuka}. This problem is sometimes referred to as the two-solar mass problem and hyperon puzzle in the literature \cite{Gandolfi0}, and is one of the hottest topics in neutron star physics. 
\par
In this paper, we pick up the neutron-star EoS, because it is a crucial key for the study of the two-solar mass problem. This is because, once it is fixed, together with the Tolman-Oppenheimer-Volkov (TOV) equation\cite{Tolman, Oppenheimer}, we can obtain the so-called $M$-$R$ relation\cite{Silbara}, linking the neutron-star mass $M$ and its radius $R$, which also gives the upper limit of the neutron-star mass. However, the determination of EoS by astronomical observations is difficult, because even the known nearest neutron star (RX J1856.5-3754) is about 400 light-years away from the earth\cite{Haberl}. Although neutron skins\cite{Tamii,Skin} and hallows\cite{Halo,Nakamura} in neutron-rich nuclei give information about neutron matter, it is still not enough to construct the neutron-star EoS, including many-body effects associated with a strong neutron-neutron interaction\cite{Dean}. As a result, the current approach to the neutron-star EoS has to strongly rely on theory\cite{Friedman,Akmal,Gezerlis,Gandolfi}. Of course, this approach is partially supported by experiment, because it employs a pseudo-potential describing neutron-neutron interaction which can reproduce few-body scattering data obtained from terrestrial experiments\cite{Friedman,Akmal,Gezerlis,Gandolfi,Wiringa}. However, since the system in question is a strongly interacting many-body system, many-body effects are expected to play important roles in a neutron-star interior. In the current approach, inclusion of these is a fully theoretical challenge. Thus, when cold Fermi gas physics can help this to some extent, it would impact on neutron star physics.
\par
%%%%%%%%%%%%%%%%%%%%%%%%%%%%%%%%%%%%%%%%%%%%%%%%%%%%%%%%%%%%%%%%%%%%%%%%%%%%%%
\par
\begin{figure}[t]
\begin{center}
\includegraphics[width=0.6\linewidth,keepaspectratio]{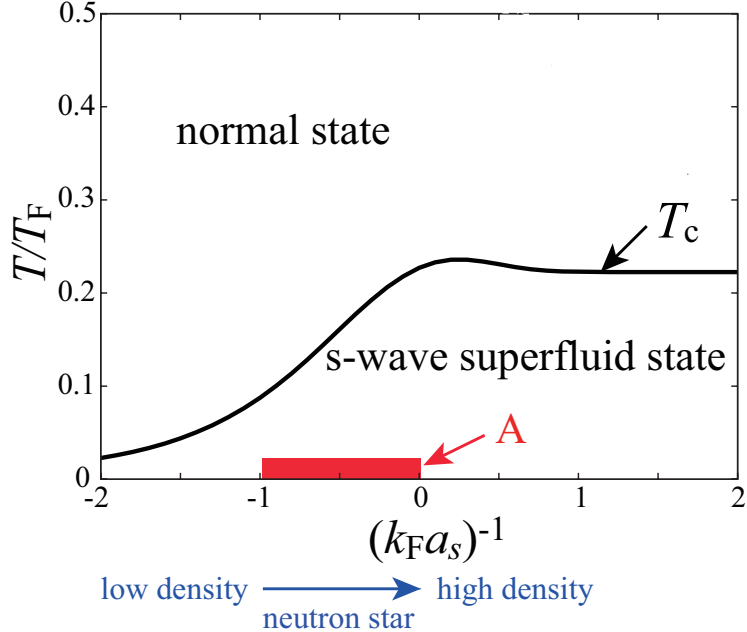}
\end{center}
\caption{(Color online) Phase diagram of an ultracold Fermi gas in the BCS-BEC crossover region. ``A" is the region where we can use for the study of neutron-star EoS in the low-density region. In this region, approaching the unitarity limit in the case of an ultracold Fermi gas corresponds to the increase of neutron density as one goes into a neutron-star interior. In this phase diagram, the interaction strength is measured in terms of the inverse $s$-wave scattering length $a_s^{-1}$, normalized by the Fermi momentum $k_{\rm F}$. The temperature is normalized by the Fermi temperature $T_{\rm F}$.}
\label{fig1}
\end{figure} 
%%%%%%%%%%%%%%%%%%%%%%%%%%%%%%%%%%%%%%%%%%%%%%%%%%%%%%%%%%%%%%%%%%%%%%%%%%%%%%%
\par
To explain our strategy, we recall the following three key issues: 
\begin{enumerate}
\item[(i)] The EoS has recently been measured with very high precision, in the BCS-unitary regime of a $^6$Li superfluid Fermi gas far below the superfluid phase transition temperature $T_{\rm c}$\cite{Horikoshi}. In this experiment, the scaled $s$-wave pairing interaction $(k_{\rm F}a_s)^{-1}$ is tuned by adjusting the $s$-wave scattering length $a_s$ by using a Feshbach resonance\cite{Chin} (where $k_{\rm F}$ is the Fermi momentum). 
\item[(ii)] In the low density regime of a neutron star interior, neutron-rich nuclei are surrounded by drip neutrons and electrons in the inner crust, and neutron matter with a small fraction of protons and electrons makes the outer core. Thus, the property of pure neutron matter is decisive in these regions. In addition, in the low density region where $n\lesssim \rho_0$ (which corresponds to the inner crust), the dominant interaction between neutrons is of an attractive s-wave type\cite{Gandolfi2}, with the scattering length $a_s=-18.5~{\rm fm}$\cite{Stoks,Howell}. Although this value is fixed in the neutron-star case, the {\it scaled interaction} $(k_{\rm F}a_s)^{-1}~(<0)$ varies to approach zero, as one goes deeper into the star. (Note that the Fermi momentum $k_{\rm F}=[3\pi^2 n]^{1/3}$ become large with increasing the density $n$). The typical magnitude $k_{\rm F}=1~{\rm fm}^{-1}$ in this regime gives ($k_{\rm F}a_s)^{-1}=-0.054$, indicating that the system is close to the unitarity limit. Since the interior temperature is considered to be much lower than the Fermi temperature $T_{\rm F}$ (except just after the birth of a neutron star), neutrons are expected to be in the strongly interacting $s$-wave superfluid state far below $T_{\rm c}$ there\cite{Dean}.
\item[(iii)] In $^6$Li and $^{40}$K Fermi atomic gases, the effective range $r_{\rm eff}$\cite{Taylor} is negligibly small, so that the scaled interaction $(k_{\rm F}a_s)^{-1}$ is the only relevant interaction parameter. However, this is not the case for interacting neutrons, where the effective range $r_{\rm eff}=2.7{\rm fm}$\cite{Slaus} cannot be ignored, because it is comparable to the typical value $k_{\rm F}^{-1}\sim 1$ fm of the inverse Fermi momentum even in the inner crust. 
\end{enumerate}
\par
Among these keys, (i) and (ii) indicate that the recent experimental achievement\cite{Horikoshi} in cold Fermi gas physics has already provided very useful information about the low density region of a neutron star interior (where the system properties are dominated by $s$-wave superfluid neutrons). The density (or radius)-dependent interaction strength $(k_{\rm F}a_s)^{-1}$ is in the latter can be simulated by the tunable interaction associated with a Feshbach resonance in the former\cite{Chin}. A crucial difference between the two is the importance of the effective range $r_{\rm eff}=2.7$ fm in the latter as mentioned in (iii). In this regard, it is difficult to modify the observed EoS data in a $^6$Li superfluid Fermi gas\cite{Horikoshi}, so as to include the non-zero effective range $r_{\rm eff}=2.7{\rm fm}$. Although there have been some theoretical investigations of the effects of the effective range on the physical properties of an ultracold Fermi gas, their experimental realization has not been achieved yet\cite{Parish,Schwenk}.
\par
In order to effectively use the similarity between (i) and (ii) overcoming the difference (iii), we take the following strategy in this paper: (1) We first deal with a superfluid Fermi gas in the BCS-unitarity limit shown as ``A" in Fig. \ref{fig1}, to theoretically explain the observed EoS in a $^6$Li superfluid Fermi gas\cite{Horikoshi} in a {\it quantitative} manner. For this purpose, we employ the strong-coupling theory developed by Nozi\`eres and Schmitt-Rink (NSR)\cite{NSR}. (2) We then extend the NSR theory so that it can treat the effective range $r_{\rm eff}$, to evaluate EoS in the low-density region of a neutron star interior in the region ``A" in Fig. \ref{fig1}.
\par
The advantage of our approach is that one can experimentally check theoretical calculations up to the inclusion of many-body strong-coupling effects (within the vanishing effective range). Thus, the ambiguity about the inclusion of many-body effects due to approximate theoretical calculations would be more suppressed than the previous approaches\cite{Friedman,Akmal,Gezerlis,Gandolfi} (where experimental support is only within few-body physics)\cite{Wiringa}. 
\par
We note that the study of quantum simulator in cold atom physics has so far mainly aimed to {\it experimentally} replicate another system, by using the high-tunability of atomic gases\cite{Bloch2,Demler}. In this sense, our approach (which uses both theory and experiment to describe a neutron star interior) is somehow different from this standard one. Regarding this, we point out that recent theoretical development in cold Fermi gas physics has enabled us to {\it quantitatively} compare calculated results with various experimental data in the BCS-BEC crossover region. Since even highly tunable cold atomic gases are still difficult to replicate all other quantum systems, it would be useful to also use this theoretical development, along with the experimental high tunability. Indeed, we will demonstrate that this combined approach gives the EoS being consistent with the previous neutron-star EoS in the low-density region.
\par
This paper is organized as follows. In Sec. II, we extend the strong-coupling NSR theory to the case with $r_{\rm eff}\ne 0$. In Sec. III, setting $r_{\rm eff}=0$, we confirm that the NSR theory can quantitatively explain the recent experiment on the internal energy $E$ in the unitary regime of a $^6$Li superfluid Fermi gas\cite{Horikoshi}. We then proceed to the case with $r_{\rm eff}\ne 0$, to examine how the EoS is affected by this quantity. Setting $r_{\rm eff}=2.7$ fm\cite{Slaus}, we calculate the neutron-star EoS in the low-density region. Throughout this paper, we set $\hbar=k_{\rm B}=1$, and the system volume $V$ is taken to be unity, for simplicity.
\par
%%%%%%%%%%%%%%%%%%%%%%%%%%%%%%%%%%%%%%%%%%%%%%%%%%%%%%%%%%%%%%%%%%%%%%%%%%%%%%
\section{Formulation}
\par
We consider a two-component uniform Fermi system, described by the Hamiltonian,
\begin{eqnarray}
H
=
\sum_{{\bm p},\sigma}\xi_{\bm p}
c^\dagger_{{\bm p},\sigma}c_{{\bm p},\sigma}
-\sum_{{\bm p},{\bm p}',{\bm q}}U({\bm p}-{\bm p}')
c^\dagger_{{\bm p}+{\bm q}/2,\up}
c^\dagger_{-{\bm p}+{\bm q}/2,\down}
c_{-{\bm p}'+{\bm q}/2,\down}
c_{{\bm p}'+{\bm q}/2,\up},
\label{eq.1} 
\end{eqnarray}   
where $c_{{\bm p},\sigma}$ is the annihilation operator of a Fermi particle with spin $\sigma=\uparrow,\downarrow$. While these are real spin states in the case of a neutron fluid, they represent pseudo-spins describing two atomic hyperfine states in an ultracold Fermi gas. In Eq. (\ref{eq.1}), $\xi_{\bm p}=\varepsilon_{\bm p}-\mu={\bm p}^2/(2m)-\mu$ is the kinetic energy of a fermion, measured from the Fermi chemical potential $\mu$, where $m$ is a particle mass. $-U({\bm p}-{\bm p}')~(<0)$ is an attractive interaction between fermions. We assume that the system is in the $s$-wave superfluid state by this pairing interaction. 
\par
In this paper, we include fluctuations in the Cooper channel within the framework of the strong-coupling theory developed by Nozi\`eres and Schmitt-Rink (NSR)\cite{NSR}, extended to the superfluid phase below $T_{\rm c}$\cite{Engelbrecht,Ohashi1,Fukushima}. For this purpose, it is convenient to divide the model Hamiltonian in Eq. (\ref{eq.1}) into the sum $H=H_{\rm MF}+H_{\rm FL}$ of the mean-field BCS part $H_{\rm MF}$ and the fluctuation part $H_{\rm FL}$. The former is written as, 
\begin{eqnarray}
H_{\rm MF}
=\sum_{{\bm p}}{\hat \Psi}^{\dagger}_{\bm p}
\left[\tilde{\xi}_{\bf p}\tau_3-\Delta_{\bm p}\tau_1\right]
{\hat \Psi}_{\bm p}
+
\sum_{\bm p}\tilde{\xi}_{\bf p}
+{1 \over 4}U({\bm 0})N_{\rm MF}^2
+
\sum_{{\bm p},{\bm p}'}
U({\bm p}-{\bm p}')
\langle 
c_{{\bm p},\uparrow}^\dagger c_{-{\bm p},\downarrow}^\dagger 
\rangle
\langle 
c_{-{\bm p}',\downarrow} c_{{\bm p}',\uparrow} 
\rangle,
\label{eq.5} 
\end{eqnarray} 
in the two-component Nambu representation\cite{Schrieffer}.
\par
Here,
\begin{eqnarray}
{\hat \Psi}_{\bm p}=
\left(
\begin{array}{c}
c_{{\bm p},\uparrow} \\
c_{-{\bm p},\downarrow}
\end{array}
\right)
\label{eq.5b}
\end{eqnarray}
is the Nambu field acting on particle-hole space, and $\tau_j~(j=1,2,3)$ are Pauli matrices. The kinetic energy ${\tilde \xi}_{\bm p}=\xi_{\bm p}-U({\bm 0})N_{\rm MF}/2$ in Eq. (\ref{eq.5}) involves the Hartree energy $-U({\bm 0})N_{\rm MF}/2$, where 
\begin{eqnarray}
N_{\rm MF}
=\sum_{{\bm p},\sigma}
\langle c_{{\bm p},\sigma}^\dagger c_{{\bm p},\sigma}\rangle
=
\sum_{\bm p}
\left[
1-{{\tilde \xi}_{\bm p} \over E_{\bm p}}\tanh{E_{\bm p} \over 2T}
\right].
\label{eq.5d}
\end{eqnarray}
The BCS superfluid order parameter,
\begin{eqnarray}
\Delta_{\bm p}
=
\sum_{{\bm p}'}U({\bm p}-{\bm p}')
\langle 
c_{{\bm p},\uparrow}^\dagger c_{-{\bm p},\downarrow}^\dagger 
\rangle
=
\sum_{{\bm p}'}U({\bm p}-{\bm p}')
{\Delta_{{\bm p}'} \over 2E_{{\bm p}'}}\tanh {E_{{\bm p}'} \over 2T},
\label{eq.5c}
\end{eqnarray}
is taken to be real and to be proportional to the $\tau_1$ component in Eq. (\ref{eq.5}), without loss of generality, where $E_{\bm p}=\sqrt{{\tilde \xi}_{\bm p}^2+\Delta_{\bm p}^2}$ describes the Bogoliubov single-particle excitations. We briefly note that the statistical average $\langle\cdot\cdot\cdot\rangle$ in Eqs. (\ref{eq.5d}) and (\ref{eq.5c}) is taken for the BCS Hamiltonian $H_{\rm MF}$ in Eq. (\ref{eq.5})\cite{Ohashi1,Fukushima}.
\par
To describe the $s$-wave superfluid state, we formally decomposed the interaction potential $U({\bm p}-{\bm p}')$ into the partial-wave components, expressing it as the sum of the $s$-wave channel ($U_s({\bm p},{\bm p}')$), $p$-wave channel ($U_p({\bm p},{\bm p}')$), $d$-wave channel ($U_d({\bm p},{\bm p}')$), and so on. Among these, only the $s$-wave channel survives in the low-momentum limit, so that one finds $U({\bm 0})=U_s({\bm 0},{\bm 0})$. Assuming that the $s$-wave interaction is the strongest in the low-density region which we are considering, we only retain this contribution in the gap equation (\ref{eq.5c}). Then, effects of the effective range $r_{\rm eff}$ can be incorporated into the theory by assuming the separable form\cite{NSR,Ho}, 
\begin{equation}
U_s({\bm p},{\bm p}')=U({\bm 0})\gamma_{\bm p}\gamma_{{\bm p}'},
\label{eq.5d2}
\end{equation}
where the basis function $\gamma_{\bm p}$ has the $s$-wave pairing symmetry, but has the following momentum dependence,
\begin{equation}
\gamma_{\bm p}=
{1 \over \sqrt{1+(p/p_{\rm c})^2}}.
\label{eq.5e}
\end{equation}
Although the choice of basis function $\gamma_{\bm p}$ in Eq. (\ref{eq.5e}) is not unique, an advantage of this choice is that the effective range theory becomes exact, when the cutoff momentum $p_{\rm c}$ is taken as
\begin{equation}
p_{\rm c}={1 \over r_{\rm eff}}
\left[
1+\sqrt{1-{2r_{\rm eff} \over a_s}}
\right]. 
\label{eq.4}
\end{equation}
(We explain the derivation of Eq. (\ref{eq.4}) in appendix A.) Here, as usual, the $s$-wave scattering length $a_s$ is related to $U({\bm 0})$ as
\begin{equation}
{4\pi a_s \over m}=
-{U({\bm 0}) \over 1-U({\bm 0})\sum_{\bm p}
{\gamma_{\bm p}^2 \over 2\varepsilon_{\bm p}}}.
\label{eq.4b}
\end{equation}
Only retaining the $s$-wave component in Eq. (\ref{eq.5d2}), we find that the superfluid order parameter $\Delta_{\bm p}$ in Eq. (\ref{eq.5c}) has the form, $\Delta_{\bm p}=\gamma_{\bm p}\Delta$, where $\Delta$ obeys
\begin{eqnarray}
1&=&U({\bm 0})\sum_{\bm p}
{\gamma_{\bm p}^2 \over 2E_{\bm p}}
\tanh{E_{\bm p} \over 2T}
\nonumber
\\
&=&
-{4\pi a_s \over m}
\sum_{\bm p}
\gamma_{\bm p}^2
\left[
{1 \over 2E_{\bm p}}
\tanh{E_{\bm p} \over 2T}
-{1 \over 2\varepsilon_{\bm p}}
\right].
\label{eq.4c}
\end{eqnarray}
In the case of a superfluid Fermi gas, where the effective range $r_{\rm eff}$ is negligibly small, one usually takes $p_{\rm c}=\infty$, or $\gamma_{\bm p}=1$ in Eq. (\ref{eq.4c}). In the neutron-star case, on the other hand, the empirical parameter set $(a_s,r_{\rm eff})=(-18.5~{\rm fm}, 2.7~{\rm fm})$ gives $p_{\rm c}=0.79~{\rm fm}^{-1}$. This implies that effects of the non-vanishing effective range become important, when the density increases to reach $k_{\rm F}\simeq p_{\rm c}\sim 1~{\rm fm}^{-1}$.
\par
Using Eq. (\ref{eq.5d2}), we can write the BCS Hamiltonian in Eq. (\ref{eq.5}) as
\begin{eqnarray}
H_{\rm MF}
=
\sum_{\bm p}
{\hat \Psi}^{\dagger}_{\bm p}
\left[\tilde{\xi}_{\bf p}\tau_3-\Delta_{\bm p}\tau_1
\right]{\hat \Psi}_{\bm p}
+
\sum_{\bm p}
\left[
\tilde{\xi}_{\bf p}+{\Delta_{\bm p}^2 \over U({\bm 0})}
\right]
+{1 \over 4}U({\bm 0})N_{\rm MF}^2.
\label{eq.5MF} 
\end{eqnarray}  
The Hamiltonian $H_{\rm FL}$ describing fluctuations in the Cooper channel is given by\cite{Ohashi1,Fukushima,Ohashi2}
\begin{equation}
H_{\rm FL}=-{U({\bm 0}) \over 2}\sum_{\bm q}
\left[
\rho_1({\bf q})\rho_1(-{\bf q})
+
\rho_2({\bf q})\rho_2(-{\bf q})
\right],
\label{eq.6} 
\end{equation} 
where
\begin{equation}
\rho_j({\bf q})=\sum_{\bm p}\gamma_{\bm p}{\hat \Psi}^\dagger_{{\bm p}+{\bm q}/2}\tau_j{\hat \Psi}_{{\bm p}-{\bm q}/2}~~(j=1,2)
\label{eq.6b}
\end{equation}
are the generalized density operators\cite{Ohashi1,Fukushima}. Since we are taking the superfluid order parameter $\Delta_{\bm p}$ being parallel to the $\tau_1$ component (see Eq. (\ref{eq.5})), $\rho_1({\bm q})$ and $\rho_2({\bm q})$ physically describe amplitude and phase fluctuations of the superfluid order parameter, respectively.
\par
We note that, in the cases of $^{40}$K and $^6$Li superfluid Fermi gases\cite{Jin,Zwierlein,Kinast,Bartenstein}, the $s$-wave pairing interaction is dominant, so that Eq. (\ref{eq.6b}) is enough to examine fluctuation corrections to system properties in the BCS-BEC crossover region. In the neutron-star case, on the other hand, non-$s$-wave interactions, such as the $p$-wave one, gradually appears with increasing the neutron density\cite{Dean}, even in the low-density region where neutrons are in the $s$-wave superfluid state. To describe this situation, one may also add corresponding fluctuation terms to $H_{\rm FL}$ in Eq. (\ref{eq.6})\cite{Ho}. However, in the current stage of cold Fermi gas physics, it is difficult to experimentally deal with such a situation. As a result, one cannot {\it experimentally} check the calculated EoS involving such non-$s$-wave strong-coupling effects. Thus, leaving the inclusion of non-$s$-wave fluctuation corrections to EoS as a future problem, we only take into account $s$-wave superfluid fluctuations described by Eq. (\ref{eq.6}) in this paper.
\par
%%%%%%%%%%%%%%%%%%%%%%%%%%%%%%%%%%%%%%%%%%%%%%%%%%%%%%%%%%%%%%%%%%%%%%%%%%%%%%
\par
\begin{figure}[t]
\begin{center}
\includegraphics[width=0.6\linewidth,keepaspectratio]{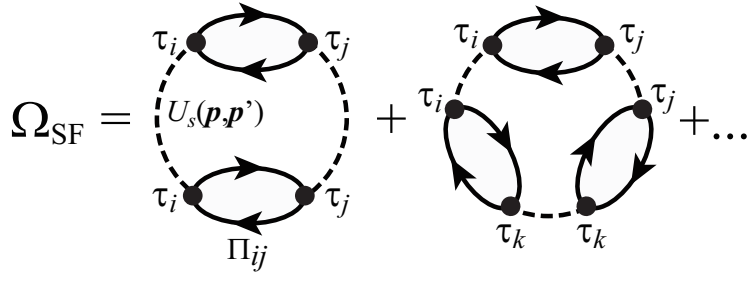}
\end{center}
\caption{Fluctuation correction $\Omega_{\rm FL}$ to the thermodynamic potential $\Omega$ in the NSR theory. The solid line and the dashed line describe the $2\times 2$ matrix single-particle BCS Green's function in Eq.(\ref{eq.12}), and the $s$-wave pairing interaction $U_s({\bm p},{\bm p}')$ in Eq. (\ref{eq.5d2}), respectively. $\Pi_{ij}$ is the pair-correlation function in Eq. (\ref{eq.11}). The solid circle is a Pauli matrix $\tau_j$.
}
\label{fig2}
\end{figure} 
%%%%%%%%%%%%%%%%%%%%%%%%%%%%%%%%%%%%%%%%%%%%%%%%%%%%%%%%%%%%%%%%%%%%%%%%%%%%%%%
\par
In the NSR theory\cite{NSR}, the thermodynamic potential $\Omega=\Omega_{\rm MF}+\Omega_{\rm FL}$ consists of the ordinary mean-field BCS part,
\begin{eqnarray}
\Omega_{\rm MF}
&=&
-T\ln\left[{\rm Tr}\left[e^{-H_{\rm MF}/T}\right]\right]
\nonumber
\\
&=&
-2T\sum_{\bm p}
\left[
\ln\left[1+e^{-E_{\bm p}/T}\right]
+
\tilde{\xi}_{\bf p}-E_{\bm p}
\right]
+{\Delta^2 \over U({\bm 0})}
+
{1 \over 4}U({\bm 0})N_{\rm MF}^2,
\label{eq.7}
\end{eqnarray}
and the fluctuation term $\Omega_{\rm FL}$ which is diagrammatically given in Fig. \ref{fig2}. Summing up these diagrams, we have
\begin{equation}
\Omega_{\rm FL}={T \over 2}\sum_{{\bs q},i\nu_n}{\rm Tr}
\Bigl[\ln\left[1+U({\bm 0})\hat{\Pi}({\bm q},i\nu_n)\right]
-U({\bm 0})\hat{\Pi}({\bm q},i\nu_n)\Bigr],
\label{eq.10}
\end{equation}
where $\nu_{n}$ is the boson Matsubara frequency. ${\hat \Pi}=\{\Pi_{ij}\}$ is the $2\times 2$ matrix pair correlation function, where \begin{equation}
\Pi_{ij}({\bm q},i\nu_n)=T\sum_{{\bm p},i\omega_n}\gamma_{\bm p}^2
{\rm Tr}\left[\tau_i{\hat G}({\bm p}+{\bm q},i\omega_n+i\nu_n)
\tau_j{\hat G}({\bm p},i\omega_n)\right].
\label{eq.11}
\end{equation}
Here,
\begin{equation}
{\hat G}({\bm p},i\omega_n)=
{1 \over
i\omega_n-{\tilde \xi}_{\bm p}\tau_3+\gamma_{\bm p}\Delta\tau_1}
\label{eq.12}
\end{equation}
is the $2\times 2$-matrix single-particle thermal Green's function in the mean-field BCS level\cite{Schrieffer}. In Eq. (\ref{eq.11}), $\Pi_{11}({\bm q},i\nu_n)$ and $\Pi_{22}({\bm q},i\nu_n)$ physically describe amplitude and phase fluctuations of the superfluid order parameter $\Delta_{\bm p}=\gamma_{\bm p}\Delta$. $\Pi_{12}({\bm q},i\nu_n)~(=-\Pi_{21}({\bm q},i\nu_n))$ represents coupling between the two fluctuations\cite{Ohashi2}.
\par
In the NSR approach, the superfluid order parameter $\Delta_{\bm p}=\gamma_{\bm p}\Delta$ and the Fermi chemical potential $\mu$ are determined by self-consistently solving the gap equation (\ref{eq.4c}), together with the equation for the total number $N$ of fermions. which is obtained from the thermodynamic identity,
\begin{equation}
N=
-\left(
{\partial \Omega \over \partial\mu}
\right)_T=N_{\rm MF}+N_{\rm FL}.
\label{eq.12b}
\end{equation}
The mean-field contribution $N_{\rm MF}=-(\partial\Omega_{\rm MF}/\partial\mu)_T$ is given in Eq. (\ref{eq.5d}). For the fluctuation correction $N_{\rm FL}=-(\partial\Omega_{\rm FL}/\partial\mu)_T$, noting that $\Omega_{\rm FL}$ depends on $\mu$ only through the effective chemical potential $\mu^*=\mu+U({\bm 0})N_{\rm MF}/2$\cite{note}, we find,
\begin{equation}
N_{\rm FL}=-
\alpha\left(
{\partial \Omega_{\rm FL} \over \partial\mu^*}
\right)_T,
\label{eq.12c}
\end{equation}
where 
\begin{eqnarray}
\alpha=
{1 \over 
1-{1 \over 2}U({\bm 0})\left({\partial N_{\rm MF} \over \partial\mu^*}\right)_T
}
\label{eq.14}
\end{eqnarray}
is the Stoner factor for the density response function\cite{Yosida}. For the derivation of Eq. (\ref{eq.12c}), see appendix B.
\par
Once $\Delta$ and $\mu$ are determined from the combined gap equation (\ref{eq.4c}) with the number equation (\ref{eq.12b}), the internal energy $E$ (or EoS) can be evaluated from $\Omega=\Omega_{\rm MF}+\Omega_{\rm FL}$, by way of the thermodynamic relation,
\begin{eqnarray}
E=
\Omega-T
\left(
{\partial \Omega \over \partial T}
\right)_\mu
-\mu
\left(
{\partial \Omega \over \partial \mu}
\right)_T.
\label{eq.15}
\end{eqnarray}
When we conveniently divide the internal energy $E=E_{\rm MF}+E_{\rm FL}$ into the the mean-field part $E_{\rm MF}$ and the fluctuations contribution $E_{\rm FL}$, each component is given by
\begin{eqnarray}
E_{\rm MF}
=
\sum_{\bm p}
\left[
E_{\bm p}f(E_{\bm p})
+{\tilde \xi}_{\bf p}-E_{\bm p}
\right]
+{\Delta^2 \over U({\bm 0})}
+
{1 \over 4}U({\bm 0})N_{\rm MF}^2+\mu N_{\rm MF},
\label{eq.15}
\end{eqnarray}
\begin{equation}
E_{\rm FL}=\Omega_{\rm FL}-
T\left({\partial\Omega_{\rm FL} \over \partial T}\right)_\mu+\mu N_{\rm FL},
\label{eq.16}
\end{equation}
where $f(x)$ is the Fermi distribution function. 
\par
The ordinary NSR formalism discussed in cold Fermi gas physics\cite{Fukushima} is immediately recovered, when we set $r_{\rm eff}\to 0$ (which leads to $p_{\rm c}\to\infty$ and $\gamma_{\bm p}\to 1$). Indeed, this limiting condition gives $U({\bm 0})\to 0$ (see Eq. (\ref{eq.4b})), so that the Stoner factor $\alpha$ in Eq. (\ref{eq.14}) is reduced to unity. In addition, the Hartree term in ${\tilde \xi}_{\bm p}=\xi_{\bm p}-U({\bm 0})N_{\rm MF}/2$, as well as the Hartree correction $U({\bm 0})N_{\rm MF}^2/4$ in Eqs. (\ref{eq.7}) and (\ref{eq.15}) vanish. Although the term $\Delta^2/U({\bm 0})$ appearing in these equations seems to diverge, this singularity is actually canceled out by the diverging behavior of the term $\sum_{\bm p}[\xi_{\bm p}-E_{\bm p}]$ in these equations, because
\begin{eqnarray}
\sum_{\bm p}\lef[\xi_{\bm p}-E_{\bm p}\rig]
+{\Delta^2 \over U({\bm 0})}
&=&
\sum_{\bm p}\lef[\xi_{\bm p}-E_{\bm p}-{\Delta^2 \over 2\varepsilon_{\bm p}}\rig]
+
{m \over 4\pi a_s}\Delta^2,
\label{eq.16b}
\end{eqnarray} 
where we have used Eq. (\ref{eq.4b}) in the first expression.
\par
Before ending this section, we comment on our numerical calculations. Although we are interested in the EoS in the ground state, we take $T/T_{\rm F}=0.01 (\ll 1)$ for computational simplicity. We briefly note that this value is much smaller than $T_{\rm c}/T_{\rm F}\sim 0.2$ in the interesting unitary regime. We have also numerically confirmed that almost the same results are obtained in the region $T/T_{\rm F}=[0.005,0.06]$. In considering a superfluid Fermi atomic gas, we set $r_{\rm eff}=0$, and the internal energy is normalized by the ground state energy $E_{\rm G}=(3/5)N\varepsilon_{\rm F}$ of a free Fermi gas, where $\varepsilon_{\rm F}$ is the Fermi energy. In the neutron-star case, we take $(a_s, r_{\rm eff})=(-18.5~{\rm fm}, 2.7~{\rm fm})$. In this case, following the convention, we measure EoS in unit of MeV, by using the neutron mass $m=936~{\rm  MeV}/c^2$ (where $c$ the speed of light).
\par
%%%%%%%%%%%%%%%%%%%%%%%%%%%%%%%%%%%%%%%%%%%%%%%%%%%%%%%%%%%%%%%%%%%%%%%%%%%%%%%
\begin{figure}[t]
\begin{center}
\includegraphics[width=0.41\linewidth,keepaspectratio]{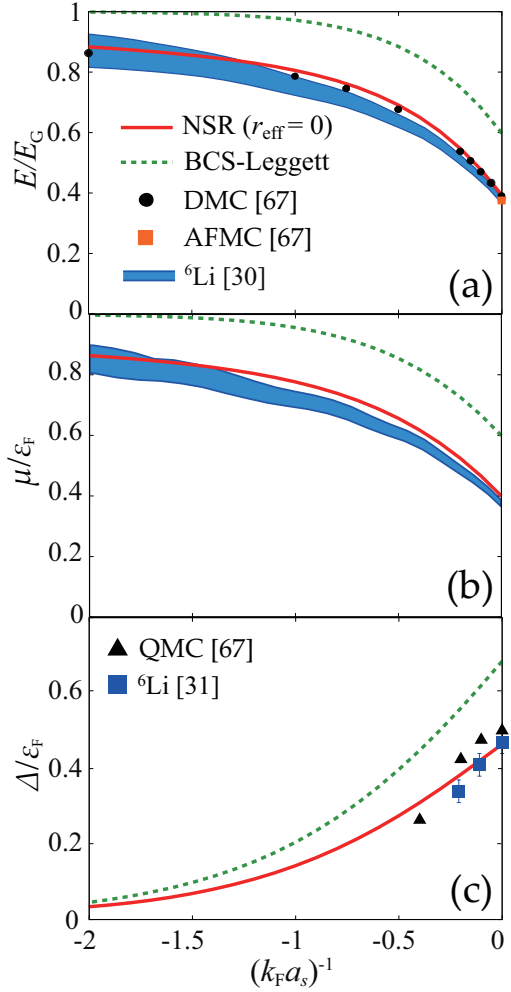}
\end{center}
\caption{(Color online) (a) Calculated internal energy $E$ in the BCS-unitary regime of a superfluid Fermi gas ($r_{\rm eff}=0$) at $T/T_{\rm F}=0.01$ (``NSR"). The dotted line shows the result in the BCS-Leggett strong-coupling theory\cite{Leggett}. ``DMC" and ``AFMC" show results by diffusion Monte-Carlo and auxiliary field Monte Carlo simulations, respectively\cite{Carlson}. The experimental result on a $^6$Li superfluid Fermi gas\cite{Horikoshi} is shown as ``$^6$Li". $E_G={2\over3}\varepsilon_{\rm F}N$ is the ground state energy of a free Fermi gas. Panels (b) and (c) show, respectively, self-consistent solutions for $\mu$ and $\Delta$, that are used in evaluating $E$ in panel (a). In panel (c), ``QMC" is the result by Monte-Carlo simulation\cite{Carlson}. ``$^6$Li" shows the experimental result by Bragg spectroscopy\cite{Vale}.
}
\label{fig3}
\end{figure} 
%%%%%%%%%%%%%%%%%%%%%%%%%%%%%%%%%%%%%%%%%%%%%%%%%%%%%%%%%%%%%%%%%%%%%%%%%%%%%%%
\par
\section{Equation of state of a neutron star in the low-density region}
\par
As mentioned previously, our approach consists of two steps, which we check one by one in this section. 
\par
\subsection{STEP 1: Assessment of the NSR theory when $r_{\rm eff}=0$}
\par
Figure \ref{fig3}(a) shows the calculated EoS, when $r_{\rm eff}=0$. While the mean-field based BCS-Leggett theory overestimates the internal energy $E$, the NSR theory well explains the recent experiment on a $^6$Li superfluid Fermi gas far below $T_{\rm c}$, as well as a Monte-Carlo simulation\cite{Carlson}. This indicates that, at least in the absence of the effective range, the NSR theory can correctly include strong-coupling corrections to the EoS, beyond the mean-field level\cite{note10}. 
\par
For completeness, we show in Figs. \ref{fig3}(b) and (c) the basic data set ($\mu$,$\Delta$) that are used in evaluating the internal energy $E$ in Fig. \ref{fig3}(a). We again find that the NSR results agree well with the recent experiments\cite{Horikoshi,Vale}, as well as a Monte-Carlo simulation\cite{Carlson}. On the other hand, the BCS-Leggett theory overestimates these quantities.
\par
%%%%%%%%%%%%%%%%%%%%%%%%%%%%%%%%%%%%%%%%%%%%%%%%%%%%%%%%%%%%%%%%%%%%%%%%%%%%%%
\begin{figure}[t]
\begin{center}
\includegraphics[width=0.56\linewidth,keepaspectratio]{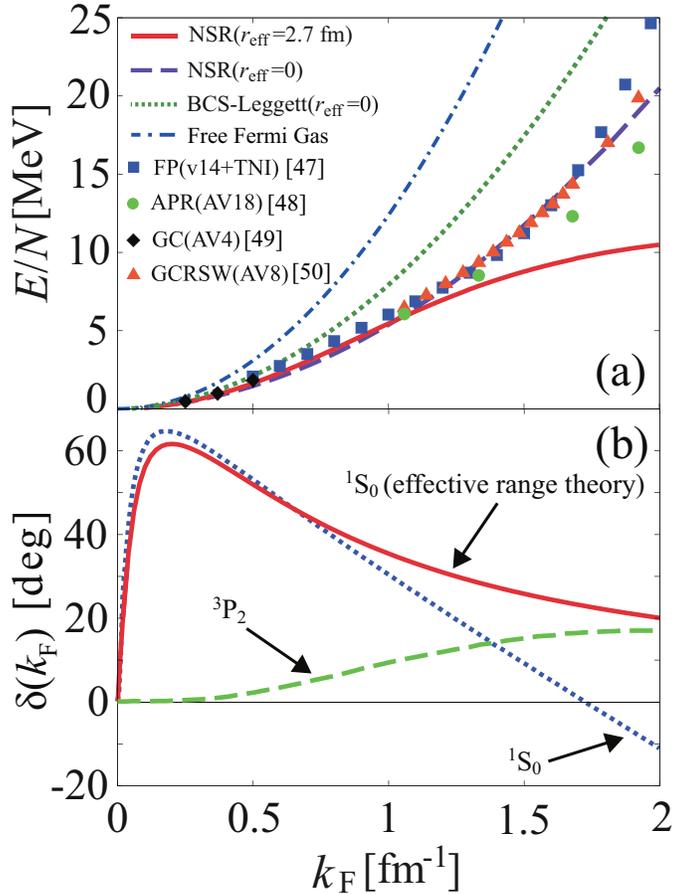}
\end{center}
\caption{(Color online) (a) Calculated equation of state (EoS) when $r_{\rm eff}=2.7$ fm (solid line). For comparison, we also show the results in the NSR theory with $r_{\rm eff}=0$ (dashed line), in the mean-field BCS-Leggett theory with $r_{\rm eff}=0$ (dotted line), as well as in a free Fermi gas (dashed-dotted line). The solid squares\cite{Friedman}, circles\cite{Akmal}, diamonds\cite{Gezerlis}, and triangles\cite{Gandolfi}, show the previous results starting from various model interactions developed in nuclear physics. (The name of the interaction is written in the parentheses.) (b) Phase shift $\delta(k_{\rm F})$ in the present $s$-wave effective range model, where the separable interaction in Eq. (\ref{eq.5d2}) with the basis function $\gamma_{\bm p}$ in Eq. (\ref{eq.5e}) is used. In this figure, we also plot the phase shift of nucleon-nucleon scattering in the $^1{\rm S}_0$ channel, as well as that in the $^3{\rm P}_2$ channel\cite{Dean,Pethick,notefig4}.}
\label{fig4}
\end{figure} 
%%%%%%%%%%%%%%%%%%%%%%%%%%%%%%%%%%%%%%%%%%%%%%%%%%%%%%%%%%%%%%%%%%%%%%%%%%%%%%%
\par
%%%%%%%%%%%%%%%%%%%%%%%%%%%%%%%%%%%%%%%%%%%%%%%%%%%%%%%%%%%%%%%%%%%%%%%%%%%%%%
\begin{figure}[t]
\begin{center}
\includegraphics[width=0.57\linewidth,keepaspectratio]{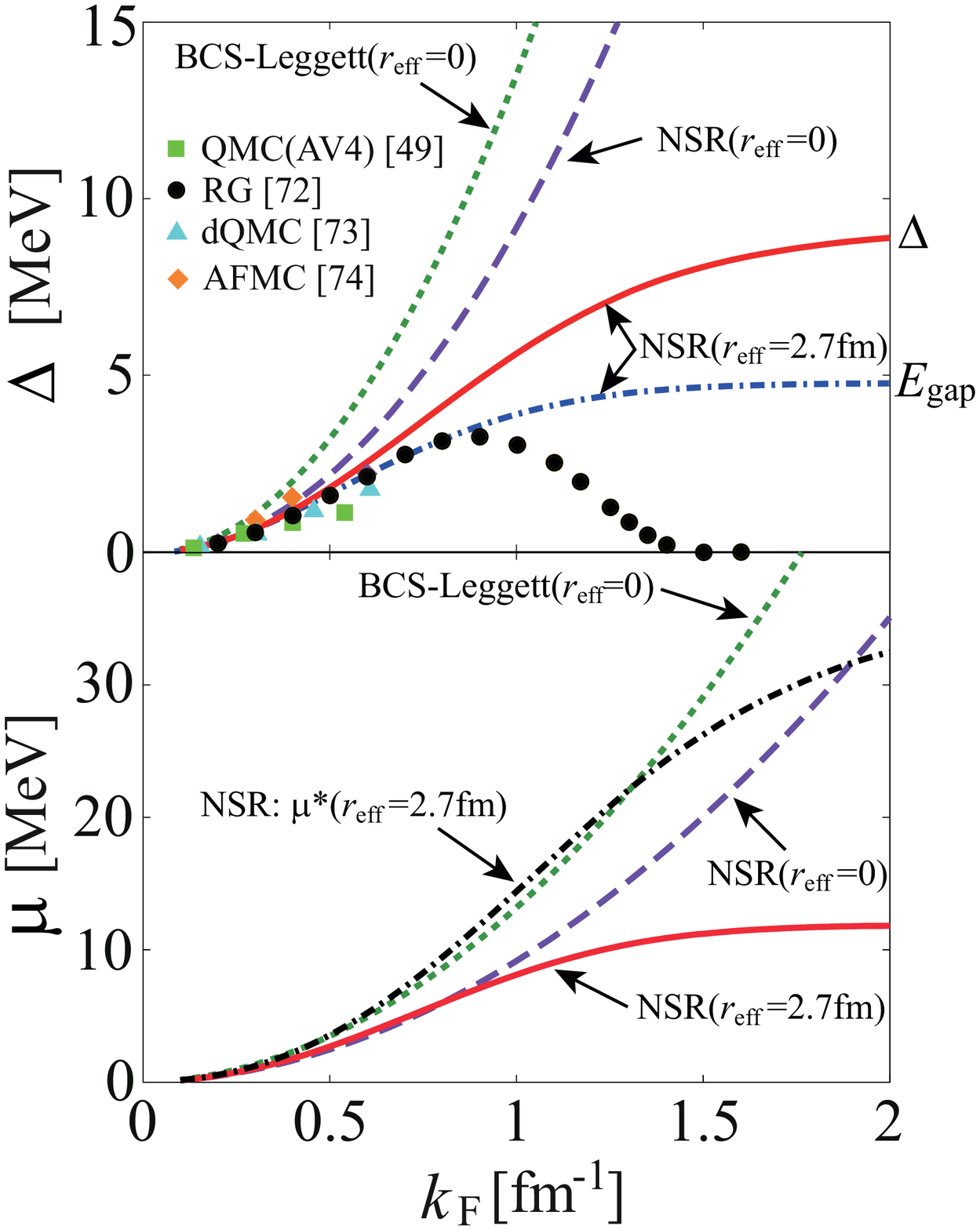}
\end{center}
\caption{(Color online) (a) Self-consistent solution for the superfluid order parameter $\Delta_{\bm p}=\gamma_{\bm p}\Delta$ when $r_{\rm eff}=2.7$ fm. $E_{\rm gap}$ is the threshold energy of the Bogoliubov single-particle dispersion $E_{\bm p}=\sqrt{{\tilde \xi}_{\bm p}^2+\Delta_{\bm p}^2}$. The dashed line and dotted line represent the superfluid order parameter $\Delta$ in the NSR theory with $r_{\rm eff}=0$ and that in the BCS-Leggett theory with $r_{\rm eff}=0$, respectively. In this figure, we also compare our result with the previous work by quantum Monte-Carlo simulation (solid squares)\cite{Gezerlis}, renormalization group (solid circles)\cite{RG}, deterministic quantum Monte-Carlo simulation (solid triangles)\cite{dQMC}, and auxiliary field Monte-Carlo simulation (solid diamonds)\cite{AFDMC2}. (b) Self-consistent solution for the chemical potential $\mu$. $\mu^*=\mu+U({\bm 0})N_{\rm MF}/2$ is the effective chemical potential. For comparison, we also plot the NSR result with $r_{\rm eff}=0$ (dashed line), as well as the result in the BCS-Leggett theory with $r_{\rm eff}=0$ (dotted line).
}
\label{fig5}
\end{figure} 
%%%%%%%%%%%%%%%%%%%%%%%%%%%%%%%%%%%%%%%%%%%%%%%%%%%%%%%%%%%%%%%%%%%%%%%%%%%%%%%
\par
\subsection{STEP 2: Application to neutron-star EoS ($r_{\rm eff}=2.7$ fm)}
\par
Building on the result in STEP 1, we now apply the same NSR theory to the case of a neutron star, by setting $r_{\rm eff}=2.7$ fm. Figure \ref{fig4}(a) shows the result, where the self-consistent solutions for $\Delta$ and $\mu$ in Fig. \ref{fig5} are used. We find that the NSR theory extended to the case with non-zero effective range well reproduces the previous results\cite{Friedman,Akmal,Gezerlis,Gandolfi} in the low density region, $k_{\rm F}\lesssim 1$ fm. As mentioned previously, although these previous calculations\cite{Friedman,Akmal,Gezerlis,Gandolfi} have used realistic neutron-neutron interactions, it has been difficult to {\it experimentally} check to what extent many-body effects are correctly taken into account in these results. In this regard, together with the result in STEP 1 (Fig. \ref{fig3}(a)), our result in Fig. \ref{fig4}(a) gives an experimental support for this point, except for effects of effective range.
\par
Figure \ref{fig4}(a) shows that our EoS gradually deviates from the previous results when $k_{\rm F}\gesim 1~{\rm fm}^{-1}$. This is simply because the effective range theory which we are using is no longer valid for such high density region. Indeed, as shown in Fig. \ref{fig4}(b), the phase shift $\delta(k_{\rm F})$ at the Fermi momentum in the effective range theory, given by
\begin{equation}
\cot\delta(k_{\rm F})=-{1 \over {k_{\rm F}a_s}}+{1 \over 2}k_{\rm F}r_{\rm eff},
\label{eq.20}
\end{equation}
gradually deviates from the $^1{\rm S}_0$ phase shift data when $k_{\rm F}\gesim p_{\rm c}=0.79~{\rm fm}^{-1}$. In addition, higher order interaction channels (e.g., $^3{\rm P}_2$ shown in Fig. \ref{fig4}(b)), as well as three-body interactions\cite{Friedman}, become important in the high-density region. While these realistic interactions are employed in the previous work\cite{Friedman,Akmal,Gezerlis,Gandolfi}, it is difficult to experimentally realize all these interactions in cold atom physics, so that our approach only deals with the already existing $s$-wave interaction.
\par
Because of the same reason, the agreement between the NSR result with $r_{\rm eff}=0$ and the previous work\cite{Friedman,Akmal,Gezerlis,Gandolfi} up to $k_{\rm F}=2~{\rm fm}^{-1}$ seen in Fig. \ref{fig4}(a) is accidental. 
\par
%%%%%%%%%%%%%%%%%%%%%%%%%%%%%%%%%%%%%%%%%%%%%%%%%%%%%%%%%%%%%%%%%%%%%%%%%%%%%%%
\par
\section{Discussions on effective-range effects from the viewpoint of $p_{\rm c}$ and $U({\bm 0})$}
\par
To understand how the effective range $r_{\rm eff}$ affects superfluid properties in more detail, it is convenient to recall that the non-vanishing $r_{\rm eff}=2.7$ fm gives a finite cutoff momentum $p_{\rm c}=0.79~{\rm fm}^{-1}$ in Eq. (\ref{eq.4}).  As a result, the region where the pairing interaction works in the gap equation (\ref{eq.4c}) is restricted to $0\le p\lesssim p_{\rm c}$. Since the region near the Fermi surface is important in the Cooper-pair formation, the growth of the superfluid order parameter $\Delta$ with increasing the Fermi momentum becomes unremarkable when $k_{\rm F}\gesim p_{\rm c}$, compared to the case of $r_{\rm eff}=0$ (giving $p_{\rm c}=\infty$). We can confirm this from the comparison of the case ``NSR($r_{\rm eff}=2.7$ fm)" with ``NSR($r_{\rm eff}=0$ fm)", as well as ``BCS-Leggett($r_{\rm eff}=0$)" in Fig. \ref{fig5}(a). 
\par
We note that the superfluid order parameter $\Delta_{\bm p}=\gamma_{\bm p}\Delta$ depends on the momentum ${\bm p}$, so that the pairing gap $E_{\rm gap}$ which is defined as the minimum excitations energy of Bogoliubov single-particle dispersion $E_{\bm p}=\sqrt{{\tilde \xi}_{\bm p}^2+\Delta_{\bm p}^2}$ does not simply equal $\Delta$, in contrast to the ordinary case with $r_{\rm eff}=0$. Indeed, the evaluated $E_{\rm gap}$ is smaller than $\Delta$ as shown in Fig. \ref{fig5}(a). This figure also shows that our result is consistent with the previous work\cite{Gezerlis,RG,dQMC,AFDMC2} in the low density region ($k_{\rm F}\lesssim 1~{\rm fm}^{-1}$).
\par
The non-vanishing effective range (or finite $p_{\rm c}$) also affects system properties through the non-zero interaction strength $U({\bm 0})$, which is related to the cutoff momentum $p_{\rm c}$ as
\begin{equation}
U({\bm 0})={4\pi a_s \over m}{1 \over 1-p_{\rm c}a_s}.
\label{eq.21}
\end{equation}
Figure \ref{fig5}(b) shows that the Fermi chemical potential is not so sensitive to the effective range, when $k_{\rm F}\lesssim 1~{\rm fm}^{-1}$. However, the so-called Hartree shift $U({\bm 0})N_{\rm MF}/2$ enlarges the effective Fermi surface size $k_{\rm F}^*\equiv \sqrt{2m\mu^*}=\sqrt{2m[\mu+U({\bm 0})N_{\rm MF}/2]}$ in this regime, which becomes comparable to the case of the BCS-Leggett theory with $r_{\rm eff}=0$ (see Fig. \ref{fig5}(b)). We briefly note that the pairing gap $E_{\rm gap}$ is obtained at the momentum which is very close to $k^*_{\rm F}$ (although we do not explicitly show the result here).
\par
We see in Fig. \ref{fig4}(a) that, while the condensation energy within the mean-field BCS-Leggett level, as well as the strong-coupling corrections within the NSR level (with $r_{\rm eff}=0$), lower the internal energy $E$, the non-vanishing effective range ($r_{\rm eff}=2.7$ fm) does not remarkably affect $E$ in the low-density region ($k_{\rm F}\lesssim 1$ fm). At a glance, this looks indicating the irrelevance of $r_{\rm eff}$ in this regime. However, Fig. \ref{fig5}(a) indicates that the effective range $r_{\rm eff}$ remarkably suppresses the superfluid order parameter $\Delta_{\bm p}=\gamma_{\bm p}\Delta$, which should also suppress the superfluid condensation energy. 
\par
The reason why we obtain $E(r_{\rm eff}=2.7~{\rm fm})\simeq E(r_{\rm eff}=0)$ in the low density region in Fig. \ref{fig4}(a) is that the above-mentioned decrease of the superfluid condensation energy is approximately compensated by the Hartree energy,
\begin{equation}
E_{\rm MF}=-{1 \over 2}U({\bm 0})N_{\rm MF}^2,
\label{eq.22}
\end{equation}
originating from the non-zero $r_{\rm eff}$. (Note that the Hartree energy vanishes when $r_{\rm eff}=0$.) This means that the mean-field Hartree energy is important in quantitatively examining the crust regime of a neutron star.
\par
%%%%%%%%%%%%%%%%%%%%%%%%%%%%%%%%%%%%%%%%%%%%%%%%%%%%%%%%%%%%%%%%%%%%%%%%%%%%%%%
\begin{figure}[t]
\begin{center}
\includegraphics[width=0.6\linewidth,keepaspectratio]{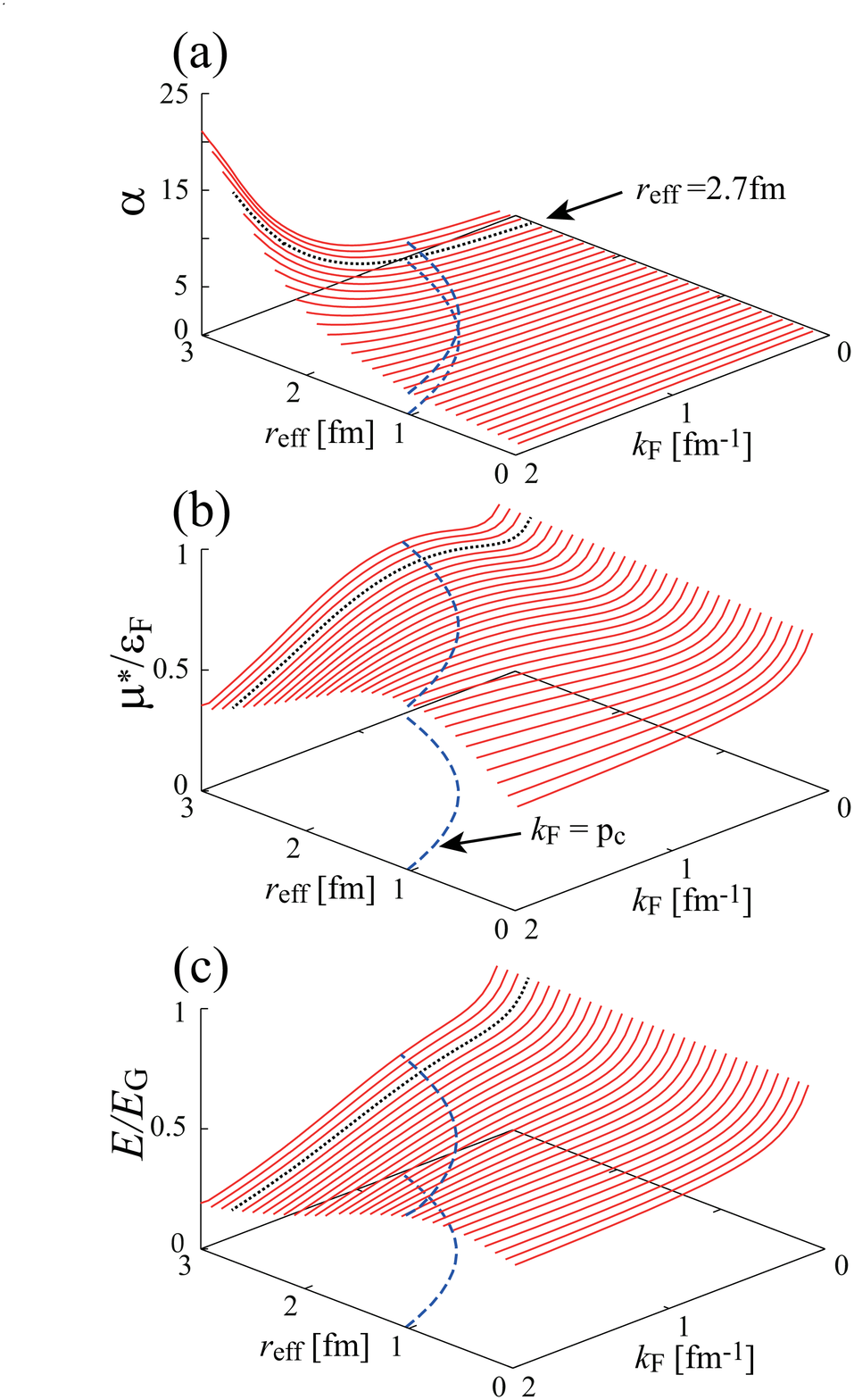}
\end{center}
\caption{(Color online) Calculated (a) Stoner factor $\alpha$ in Eq. (\ref{eq.14}), (b) effective chemical potential $\mu^*=\mu+U({\bm 0})N_{\rm MF}/2$, and (c) internal energy $E$, as functions of the Fermi momentum $k_{\rm F}$ and the effective range $r_{\rm eff}$. We take $a_s=-18.5$ fm. In each panel, the dotted line and the dashed line show the result at $r_{\rm eff}=2.7$ fm, and that at $k_{\rm F}=p_{\rm c}=0.79~{\rm fm}^{-1}$, respectively.}
\label{fig6}
\end{figure} 
%%%%%%%%%%%%%%%%%%%%%%%%%%%%%%%%%%%%%%%%%%%%%%%%%%%%%%%%%%%%%%%%%%%%%%%%%%%%%%%
\par
Before ending this section, we comment on two other effects associated with the effective range $r_{\rm eff}$. First, the non-zero $U({\bm 0})$ produces the Stoner factor $\alpha$ in Eq. (\ref{eq.14}), which enhances the NSR fluctuation contribution $N_{\rm FL}$ to the number equation in Eq. (\ref{eq.12c}). However, we see in Fig. \ref{fig6} that the region where the effective chemical potential $\mu^*$, as well as the internal energy $E$, are strongly influenced by the Stoner enhancement is restricted to the high-density region $k_{\rm F}\gesim 1~{\rm fm}^{-1}$. Thus, as far as we consider the low-density region ($k_{\rm F}\lesssim 1~{\rm fm}^{-1}$), this effective-range effect does not seem important.
\par
Second, when $r_{\rm eff}=0$, the magnitude of each diagram in Fig. \ref{fig2} is not well-defined, because $U({\bm 0})=+0$ and the pair correlation function $\Pi_{ij}$ in Eq. (\ref{eq.11}) exhibits the ultraviolet divergence. Their infinite summation only gives a finite fluctuation correction $\Omega_{\rm FL}$ to the thermodynamic potential $\Omega$. In contrast, when $r_{\rm eff}>0$, each diagram in Fig. \ref{fig2}, as well as the other diagrams that are ignored in the NSR theory, become non-zero because of $U({\bm 0})>0$. In this case, since the superfluid order is weakened by the effective range (see Fig. \ref{fig5}(a)), it becomes unclear whether the NSR scheme (where special diagrams describing superfluid fluctuations are selectively summed up to the infinite order) is still superior to the perturbative order-by-order calculation in terms of the pairing interaction. Regarding this, explicitly evaluating {\it all} the second-order diagrams contributing to the thermodynamic potential that are not taken into account in the NSR theory, we find that the correction ($\equiv E_{\rm corr}$) to the EoS is very small, as shown in Fig. \ref{fig7}. (For the derivation of $E_{\rm corr}$, see appendix C.) This means that the inclusion of superfluid fluctuations described by the diagrammatic series in Fig. \ref{fig2} is still effective in considering the low-density region of a neutron-star interior.
\par
%%%%%%%%%%%%%%%%%%%%%%%%%%%%%%%%%%%%%%%%%%%%%%%%%%%%%%%%%%%%%%%%%%%%%%%%%%%%%%%
\begin{figure}[t]
\begin{center}
\includegraphics[width=0.6\linewidth,keepaspectratio]{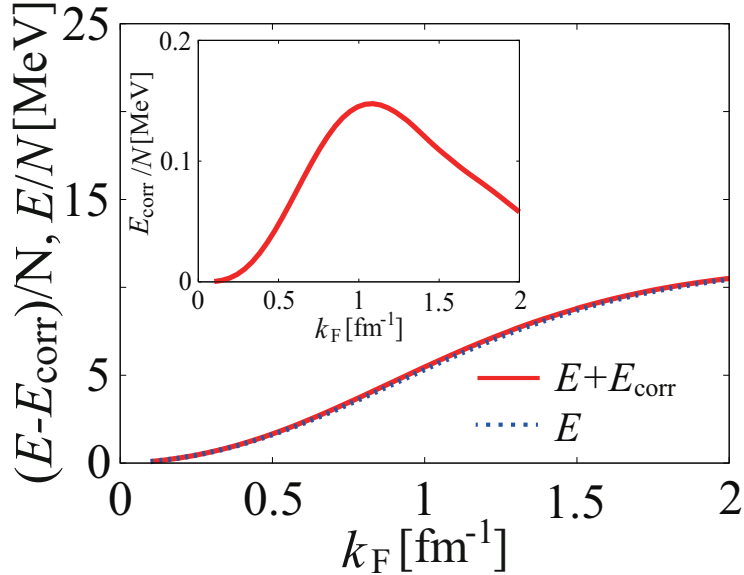}
\end{center}
\caption{(Color online) Internal energy $E+E_{\rm corr}$, including both the NSR contribution and the second-order correction ($E_{\rm corr}$) that are ignored in the NSR scheme (see the inset). $E$ is the internal energy in the NSR theory (when $r_{\rm eff}=2.7$ fm).}
\label{fig7}
\end{figure} 
%%%%%%%%%%%%%%%%%%%%%%%%%%%%%%%%%%%%%%%%%%%%%%%%%%%%%%%%%%%%%%%%%%%%%%%%%%%%%%%
\par
\section{Summary}
\par
To summarize, we have discussed a possible application of an ultracold Fermi atomic gas to the study of a neutron star equation-of-state (EoS). Although our idea maximally uses the high-tunability of this atomic system, we do not attempt to experimentally replicate a neutron star by using the high tunability of an ultracold Fermi gas, but simply use the already existing superfluid state. That is, noting that the inner crust regime of a neutron star is considered to be in the nearly unitary $s$-wave superfluid state of neutrons far below $T_{\rm c}$, we first deal with the recent experiment on EoS in a superfluid $^6$Li Fermi gas in the BCS-unitary regime\cite{Horikoshi}. We then theoretically make up for the crucial difference between the two systems about the magnitude of the effective range $r_{\rm eff}$, because it cannot experimentally be tuned in the current stage of cold atom physics.
\par
To demonstrate our idea, we first showed that the recent EoS measurement on a $^6$Li superfluid Fermi gas can be quantitatively explained by the strong-coupling theory developed by Nozi\`eres and Schmitt-Rink (NSR). We then extended the NSR theory to include the non-vanishing effective range ($r_{\rm eff}=2.7$ fm), so as to be able to treat the inner crust regime of a neutron star. The calculated EoS was found to agree well with the previous theoretical work on the neutron-star EoS in this regime. Although these previous calculations use detailed neutron-neutron interactions which can reproduce the experimental phase shift data, no experimental support has existed about the inclusion of many-body effects associated with strong pairing interaction near the unitarity limit. Our combined strong-coupling theory with cold-Fermi-gas experiment gives confirmation about this  for the first time, except for effects of the non-zero effective range. 
\par
Since the present approach is only valid for the low-density region ($k_{\rm F}\lesssim 1~{\rm fm}^{-1}$) of a neutron star, it is an exciting challenge to extend this to the deeper core region, where the simple $s$-wave neutron superfluid is no longer expected. In this regard, one possibility is to use a $p$-wave superfluid Fermi gas. At present, while a tunable $p$-wave pairing interaction associated with a $p$-wave Feshbach resonance\cite{Tichnor,Nakatsuji}, as well as the formation of $p$-wave pairs\cite{Zhang,Jin2003}, have been realized, any $p$-wave superfluid state has not been achieved yet, because of very short lifetime of $p$-wave pairs\cite{Chevy} due to three-body loss\cite{Castin,Levinsen}, as well as dipolar relaxation\cite{Jin2007}. However, once a $p$-wave superfluid Fermi atomic gas is realized, we would be able to use it as a testing ground, to construct a strong-coupling theory which can {\it quantitatively} describe a $p$-wave Fermi superfluid. Even if the detailed $p$-wave pairing symmetry in the case of an ultracold Fermi gas is different from that expected in the core region of a neutron star, the strong-coupling theory which is experimentally assessed in the former would be useful for the study of the core region where a $p$-wave neutron superfluid is expected (see Fig. \ref{fig4}(a)), by modifying the theory to compensate the difference between the two systems (as we have done in the $s$-wave case). Such an application would also be a good motivation for the research toward the realization of a $p$-wave superfluid Fermi gas. Since it is difficult to directly measure the neutron-star interior, our idea would provide an alternative route to this astronomical object, in addition to the conventional approach being based on nuclear physics.
\par
%%%%%%%%%%%%%%%%%%%%%%%%%%%%%%%%%%%%%%%%%%%%%%%%%%%%%%%%%%%%%%%%%%%%%%%%%%%%%%%
\par
\acknowledgements
We thank M. Matsuo, T. Tatsumi, T. Takatsuka, M. Horikoshi, R. Hanai, M. Matsumoto for useful discussions. This work was supported by KiPAS project at Keio University. H.T. was supported by a Grant-in-Aid for JSPS fellows. Y.O. was supported by Grant-in-Aid for Scientific research from MEXT and JSPS in Japan (No.16K05503, No.15K00178, No.15H00840).
\par
%%%%%%%%%%%%%%%%%%%%%%%%%%%%%%%%%%%%%%%%%%%%%%%%%%%%%%%%%%%%%%%%%%%%%%%%%%%%%%%
\par
\appendix
\par
\section{Effective range theory in the case of the basis function $\gamma_{\bm p}$ in Eq. (\ref{eq.5e})}
\par
We consider a two-particle system with the separable $s$-wave interaction in Eq. (\ref{eq.5d2}). The two-particle scattering $T$-matrix $\Gamma_s({\bm p},{\bm p}',\omega_+)$ obeys\cite{Ho},
\begin{eqnarray}
\Gamma_s({\bm p},{\bm p}';\omega_+)
=
-U_s({\bm p},{\bm p}')
-
\sum_{\bm k}U_s({\bm p},{\bm k})
{1 \over \omega_+-2\varepsilon_{\bm k}}
\Gamma_s({\bm k},{\bm p'};\omega_+),
\label{eq.a1}
\end{eqnarray}
where $\omega_+=\omega+i\delta$, with $\delta$ being an infinitesimally small positive number. Equation (\ref{eq.a1}) gives $\Gamma_s({\bm p},{\bm p}';\omega_+)=\gamma_{\bm p}\Lambda_s(\omega_+)\gamma_{{\bm p}'}$, where 
\begin{equation}
{1 \over \Lambda_s(\omega_+)}=
-{1 \over U({\bm 0})}
-\sum_{\bm p}{\gamma_{\bm p}^2 \over \omega_+-2\varepsilon_{\bm p}}.
\label{eq.a2}
\end{equation}
The scattering $T$-matrix $\Gamma({\bm p},{\bm p}';\omega_+)$ is related to the scattering amplitude $f_s({\bm p})$ as\cite{Taylor},
\begin{equation}
f_s({\bm p})=-{m \over 4\pi}\Gamma_s({\bm p},{\bm p};2\varepsilon_{\bm p}+i\delta).
\label{eq.a3}
\end{equation}
Using Eqs. (\ref{eq.4b}) and (\ref{eq.a2}), one finds that the scattering amplitude $f({\bm p})$ in Eq. (\ref{eq.a3}) is written as
\begin{equation}
f_s({\bm p})=
{\gamma_{\bm p}^2 \over \displaystyle
-{1 \over a_s}-{4\pi \over m}\sum_{{\bm p}'}\gamma_{{\bm p}'}^2
\left[
{1 \over 2\varepsilon_{{\bm p}'}-(2\varepsilon_{\bm p}+i\delta)}
-{1 \over 2\varepsilon_{{\bm p}'}}
\right]}.
\label{eq.a4}
\end{equation}
When we take the basis function $\gamma_{\bm p}$ in Eq. (\ref{eq.5e}) (where the cutoff momentum $p_{\rm c}$ is given in Eq. (\ref{eq.4})), Eq. (\ref{eq.a4}) gives the {\it exact} expression in the effective range theory\cite{Taylor},
\begin{equation}
f_s({\bm p})={1 \over \displaystyle 
-{1 \over a_s} +{1 \over 2}r_{\rm eff}p^2-ip}.
\label{eq.a5}
\end{equation}
We briefly note that, higher order terms (such as $\sim p^4$) generally appears in the denominator of Eq. (\ref{eq.a5}), when one chooses another expression for $\gamma_{\bm p}$, e.g., $\gamma_{\bm p}=1/[1+(p/p_{\rm c})^2]$.
\par
%%%%%%%%%%%%%%%%%%%%%%%%%%%%%%%%%%%%%%%%%%%%%%%%%%%%%%%%%%%%%%%%%%%%%%%%%%%%%%%
\par
\section{Derivation of Eq. (\ref{eq.12c}) and how to evaluate Eq. (\ref{eq.14})}\par
Noting that $\Omega_{\rm FL}$ depends on $\mu$ only through $\mu^*=\mu+U({\bm 0})N_{\rm MF}/2$, one finds,
\begin{equation}
N_{\rm FL}=-
\left(
{\partial \Omega_{\rm FL} \over \partial\mu}
\right)_T=
\left(
{\partial \Omega_{\rm FL} \over \partial\mu^*}
\right)_T
\left(
{\partial \mu^* \over \partial\mu}
\right)_T,
\label{eq.b1}
\end{equation}
where
\begin{eqnarray}
\left(
{\partial \mu^* \over \partial\mu}
\right)_T
&=&
1+{1 \over 2}U({\bm 0})
\left(
{\partial N_{\rm MF} \over \partial\mu^*}
\right)_T
\left(
{\partial \mu^* \over \partial\mu}
\right)_T
\nonumber
\\
&=&
{1 \over 
1-{1 \over 2}U({\bm 0})\left(
{\partial N_{\rm MF} \over \partial\mu^*}\right)_T
},
\label{eq.b2}
\end{eqnarray}
which just equals the Stoner factor $\alpha$ in Eq. (\ref{eq.14}). In Eq. (\ref{eq.b2}), we have used the fact that $N_{\rm MF}$ depends on $\mu$ only through $\mu^*$ (see Eqs. (\ref{eq.5d}) and (\ref{eq.4c})).
\par
To evaluate the factor $(\partial N_{\rm MF}/\partial\mu^*)_T$ in Eq. (\ref{eq.b2}), we conveniently abbreviate the right hand side of Eq. (\ref{eq.5d}) as $g_N(\mu^*,\Delta(\mu^*),T)$, and that of Eq. (\ref{eq.4c}) as $g_\Delta(\mu^*,\Delta(\mu^*),T)$. From Eq. (\ref{eq.4c}), we find
\begin{equation}
\left(
{\partial g_\Delta \over \partial\mu^*}
\right)_{T}
=
\left(
{\partial g_\Delta \over \partial\mu^*}
\right)_{\Delta,T}+
\left(
{\partial g_\Delta \over \partial\Delta}
\right)_{\mu^*,T}
\left(
{\partial \Delta \over \partial\mu^*}
\right)_T=0.
\label{eq.b3}
\end{equation}
Taking the partial derivative of Eq. (\ref{eq.5d}) with respect to $\mu^*$, one obtains
\begin{eqnarray}
\left(
{\partial N_{\rm MF} \over \partial\mu^*}
\right)_T
&=&
\left(
{\partial g_N \over \partial\mu^*}
\right)_{\Delta,T}
+
\left(
{\partial g_N \over \partial\Delta}
\right)_{\mu^*,T}
\left(
{\partial \Delta \over \partial\mu^*}
\right)_T
\nonumber
\\
&=&
\left(
{\partial g_N \over \partial\mu^*}
\right)_{\Delta,T}
-
\left(
{\partial g_N \over \partial\Delta}
\right)_{\mu^*,T}
\left(
{\partial g_\Delta \over \partial\mu^*}
\right)_{\Delta,T}
\left(
{\partial g_\Delta \over \partial\Delta}
\right)_{\mu^*,T}^{-1}.
\nonumber
\\
\label{eq.b4}
\end{eqnarray}
\par
%%%%%%%%%%%%%%%%%%%%%%%%%%%%%%%%%%%%%%%%%%%%%%%%%%%%%%%%%%%%%%%%%%%%%%%%%%%%%%%
\par
\section{Second-order correction $\Omega_{\rm corr}$ to thermodynamic potential}\par
To evaluate all the second-order corrections to the thermodynamic potential in a systematic manner, we conveniently note that the interaction part $H_{\rm FL}$ of the Hamiltonian in Eq. (\ref{eq.6}) can be written in the the following two forms.
\begin{equation}
H_{\rm FL}=-U({\bm 0})
\sum_{{\bm p},{\bm p}',{\bm q}}
\gamma_{\bm p}\gamma_{\bm p'}
\rho_+({\bm p},{\bm q})
\rho_-({\bm p}',-{\bm q}),
\label{eq.c1} 
\end{equation}
\begin{equation}
H_{\rm FL}=-U({\bm 0})
\sum_{{\bm p},{\bm p}',{\bm q}}
\gamma_{({\bm p}+{\bm p}'+{\bm q})/2}
\gamma_{({\bm p}+{\bs p}'-{\bm q})/2}
n_+({\bm p},{\bm q})
n_-({\bm p}',-{\bm q}),
\label{eq.c2} 
\end{equation}
where
\begin{equation}
\rho_{\pm}({\bm p},{\bm q})=
{1 \over 2}
\Psi^\dagger_{{\bm p}+{\bm q}/2}
[\tau_1\pm i\tau_2]
\Psi_{{\bm p}-{\bm q}/2},
\label{eq.c3}
\end{equation}
\begin{equation}
n_\pm({\bm p},{\bm q})=
{1 \over 2}
\Psi^\dagger_{{\bm p}+{\bm q}/2}
[\tau_3\pm\tau_0]
\Psi_{{\bm p}-{\bm q}/2},
\label{eq.c4}
\end{equation}
with $\tau_0$ being the unit matrix. Physically, $\rho_\pm({\bm p},{\bm q})$ and $n_\pm({\bm p},{\bm q})$ describe superfluid fluctuations and density fluctuations, respectively.
\par
The expression for the second order correction $\Omega_{\rm corr}$ to the thermodynamic potential in terms of $H_{\rm FL}$ is obtained by using the linked cluster theorem\cite{AGD} as,
\begin{equation}
\Omega_{\rm corr}=-{T \over 2}
\int_0^{1/T}{\rm d}\tau\int_0^{1/T}{\rm d}\tau'
\langle H_{\rm FL}(\tau)H_{\rm FL}(\tau')\rangle_{\rm c}.
\label{eq.c5} 
\end{equation}
Here, $H_{\rm FL}(\tau)=e^{H_{\rm MF}\tau}H_{\rm FL}e^{-H_{\rm MF}\tau}$, and $\langle\cdot\cdot\cdot\rangle_{\rm c}$ only involves contributions from connected diagrams. When one uses Eq. (\ref{eq.c1}) for the two $H_{\rm FL}$'s in Eq. (\ref{eq.c5}), the result is the same as that obtained from the second-order diagram in Fig. \ref{fig2} ($\equiv \Omega_{\rm FL}^{(2)}$), which has, of course, already been included in $\Omega_{\rm FL}$ in Eq. (\ref{eq.10}). The second-order correction $\Omega_{\rm FL}^{(2)}$ is also reproduced, when one uses Eq. (\ref{eq.c2}) for the two $H_{\rm FL}$'s in Eq. (\ref{eq.c5}). This is because, although the second-order diagram in Fig. \ref{fig2} is treated as that describing superfluid fluctuations in the NSR theory, it may actually be regarded as a diagram describing fluctuations in the density channel. As a result, we should also drop this contribution, to avoid double-counting.
\par
The second-order correction which is not involved in the NSR theory is obtained when one uses Eq. (\ref{eq.c1}) for one of the two $H_{\rm FL}$'s and Eq. (\ref{eq.c2}) for the other $H_{\rm FL}$ in Eq. (\ref{eq.c5}), which gives
\begin{eqnarray}
\Omega_{\rm corr}
&=&
-U({\bm 0})^2
T\sum_{{\bm p},{\bm p}',{\bm q},\nu_n}
\gamma_{({\bm p}+{\bm p}'+{\bm q})/2}
\gamma_{({\bm p}+{\bm p}'-{\bm q})/2}
\gamma_{\bm p}
\gamma_{{\bm p}'}
\nonumber
\\
&\times&
\Bigl[
\Pi^{\rho n}_{++}({\bm p},{\bm q},i\nu_n)
\Pi^{n\rho}_{--}({\bm p}',-{\bm q},i\nu_n)
+
\Pi^{\rho n}_{+-}({\bm p},{\bm q},i\nu_n)
\Pi^{n\rho}_{+-}({\bm p}',-{\bm q},i\nu_n)
\Bigr]
\nonumber
\\
&=&
-2U({\bm 0})^2T	
\sum_{{\bm p},{\bm p}',{\bm q},\nu_n}
\gamma_{({\bm p}+{\bm p}'+{\bm q})/2}
\gamma_{({\bm p}+{\bm p}'-{\bm q})/2}
\gamma_{\bm p}
\gamma_{{\bm p}'}
\Pi^{\rho n}_{++}({\bm p},{\bm q},i\nu_n)
\Pi^{\rho n}_{++}({\bm p}',{\bm q},i\nu_n),
\label{eq.c8}
\end{eqnarray}
where
\begin{equation}
\Pi^{\rho n}_{ij}({\bm p},{\bm q},i\nu_n)=T\sum_{\nu_n}
{\rm tr}
\left[
\tau_i
\hat{G}({\bm p}+{\bm q}/2,i\omega_n+i\nu_n)
n_j
\hat{G}({\bm p}-{\bm q}/2,i\omega_n)
\right],
\label{eq.c7}
\end{equation}
\begin{equation}
\Pi^{n\rho}_{ij}({\bm p},{\bm q},i\nu_n)=T\sum_{\nu_n}
{\rm tr}
\left[
n_i
\hat{G}({\bm p}+{\bm q}/2,i\omega_n+i\nu_n)
\tau_j
\hat{G}({\bm p}-{\bm q}/2,i\omega_n)
\right],
\label{eq.c7b}
\end{equation}
physically describe couplings between superfluid fluctuations and density fluctuations\cite{Ohashi2}. In obtaining the last expression in Eq. (\ref{eq.c8}), we have used the symmetry properties, $\Pi^{n\rho}_{--}({\bm p},-{\bm q},i\nu_n)=\Pi^{\rho n}_{+-}({\bm p},-{\bm q},i\nu_n)=\Pi^{\rho n}_{++}({\bm p},{\bm q},i\nu_n)$, and $\Pi^{n\rho}_{+-}({\bm p},{\bm q},i\nu_n))=\Pi^{\rho n}_{++}({\bm p},{\bm q},i\nu_n)$. Summing up the Matsubara frequencies in $\Pi_{++}^{\rho n}$ in Eq. (\ref{eq.c8}), we have
\begin{eqnarray}
\Pi^{\rho n}_{++}({\bm p},{\bm q},i\nu_n))&=&-{\Delta_{{\bm p}+{\bm q}/2}\over 4E_{{\bm p}+{\bm q}/2}}
\Bigg[
\left(1+{\tilde{\xi}_{{\bm p}-{\bm q}/2} \over E_{{\bm p}-{\bm q}/2}}\right)
\left[
{
1 - f(E_{{\bm p}+{\bm q}/2})-f(E_{{\bm p}-{\bm q}/2}) \over 
i\nu_n+E_{{\bm p}+{\bm q}/2}+E_{{\bm p}-{\bm q}/2}
}
-
{f(E_{{\bm p}+{\bm q}/2})-f(E_{{\bm p}-{\bm q}/2}) \over {i\nu_n-E_{{\bm p}+{\bm q}/2}}+E_{{\bm p}-{\bm q}/2}}
\right]
\nonumber\\
&+&\left(1-{\tilde{\xi}_{{\bm p}-{\bm q}/2} \over E_{{\bm p}-{\bm q}/2}}\right)
\left[
{1 - f(E_{{\bm p}+{\bm q}/2})-f(E_{{\bm p}-{\bm q}/2})  \over i\nu_n-E_{{\bm p}+{\bm q}/2}-E_{{\bm p}-{\bm q}/2}}
-
{f(E_{{\bm p}+{\bm q}/2})-f(E_{{\bm p}-{\bm q}/2}) \over i\nu_n+E_{{\bm p}+{\bm q}/2}-E_{{\bm p}-{\bm q}/2}}
\right]
\Bigg].
\label{eq.c9}
\end{eqnarray}
Substituting Eq. (\ref{eq.c9}) into Eq. (\ref{eq.c8}), which is followed by the $\nu_n$-summation, we obtain, in the low temperature limit,
\begin{equation}
\Omega_{\rm corr}={U({\bm 0})^2 \over 4}
\sum_{{\bm p}, {\bm p}',{\bm q}}
\left[
1-{\tilde{\xi}_{{\bm p}-{\bm q}/2} \over E_{{\bm p}-{\bm q}/2}}
\right]
\left[
1-{\tilde{\xi}_{{\bm p}'-{\bm q}/2} \over E_{{\bm p'}-{\bm q}/2}}
\right]
{
\gamma_{\bm p}
\gamma_{{\bm p}'}
\gamma_{({\bm p}+{\bm p}'+{\bm q})/2}
\gamma_{({\bm p}+{\bm p}'-{\bm q})/2}
\Delta_{{\bm p}+{\bm q}/2}\Delta_{{\bm p}'+{\bm q}/2} 
\over 
E_{{\bm p}+{\bm q}/2}E_{{\bm p}'+{\bm q}/2}
\left[E_{{\bm p}+{\bm q}/2}+E_{{\bm p}-{\bm q}/2}+E_{{\bm p}'+{\bm q}/2}+E_{{\bm p'}-{\bm q}/2}\rig]}
.
\label{eq.17}
\end{equation}
To obtain Fig. \ref{fig7}, we have numerically solved the gap equation (\ref{eq.4c}), together with the {\it modified} number equation $N=N_{\rm MF}+N_{\rm FL}+N_{\rm corr}$, where
\begin{equation}
N_{\rm corr}=-\alpha
\left(
{\partial\Omega_{\rm corr}} \over {\partial\mu^{*}}
\right)_T.
\label{eq.18}
\end{equation}                    						
The correction $E_{\rm corr}$ to the internal energy is calculated from 
\begin{equation}
E_{\rm corr}=\Omega_{\rm corr}-T\lef({\partial\Omega_{\rm corr}}\over{\partial T}\rig)_{\mu}+\mu N_{\rm corr}.
\label{eq.19}
\end{equation}
\par
%%%%%%%%%%%%%%%%%%%%%%%%%%%%%%%%%%%%%%%%%%%%%%%%%%%%%%%%%%%%%%%%%%%%%%%%%%%%%%%
\par


\begin{thebibliography}{100}
\bibitem{note} The gap equation (\ref{eq.4c}) indicates that the superfluid order parameter $\Delta$ depends on $\mu$ only through $\mu^*$. Thus, the fluctuation part $\Omega_{\rm FL}$ of the thermodynamic potential may be written as $\Omega_{\rm FL}(T,\mu^*,\Delta(\mu^*))=\Omega_{\rm FL}(T,\mu^*)$.
\bibitem{note10} It has been shown\cite{Tajima} that the slight differences between the NSR and the experimental results seen in Figs. \ref{fig3}(a) and (b) are removed by including higher order pairing fluctuations beyond the NSR theory. However, leaving this extension as a future problem, we examine effects of non-vanishing effective range $r_{\rm eff}$ within the NSR scheme in this paper.
\bibitem{notefig4} Strictly speaking, the $^1{\rm S}_0$ and $^3{\rm P}_2$ phase shifts are for neutron-proton interactions\cite{Pethick}. However, since the isospin symmetry is only weakly broken, they approximately describe the neutron-neutron case. We note that the slight different peak heights between our model phase shift and the neutron-proton case in Fig. \ref{fig4}(b) is due to the stronger neutron-proton interaction $a_s=-23.5~{\rm fm}$ than the neutron-neutron case $a_{s}=-18.5~{\rm fm}$.
\end{thebibliography}
\end{document}